\documentclass[12pt,a4paper,final]{iopart}

\usepackage[breaklinks=true, colorlinks=true, linkcolor=blue, urlcolor=blue, citecolor=blue]{hyperref}
\usepackage{amsfonts, amsmath, dsfont}
\usepackage{cite}

\def\epp{\: .}
\def\epc{\: ,}
\newcommand{\sine}[2]{s_{#1,#2}}

\begin{document}

\title[Overlaps of $q$-raised N\'eel states with XXZ Bethe states]{Overlaps of $q$-raised N\'eel states with XXZ Bethe states and their relation to the Lieb-Liniger Bose gas}

\author{M. Brockmann
}
\address{
Institute for Theoretical Physics, University of Amsterdam, Science Park 904,\\
Postbus 94485, 1090 GL Amsterdam, The Netherlands}
\ead{M.Brockmann@uva.nl}

\begin{abstract}
We present a `Gaudin-like' determinant expression for overlaps of $q$-raised N\'eel states with Bethe states of the spin-1/2 XXZ chain in the non-zero magnetization sector. The former is constructed by applying global $U_q(sl_2)$ spin raising operators to the N\'eel state, the ground state of the antiferromagnetic Ising chain.\\ 
The formulas presented are derived from recently-obtained results for the overlap of the N\'eel state with XXZ Bethe states \cite{XXZpaper, Pozsgay_1309.4593, 2012_Kozlowski_JSTAT_P05021, 1998_Tsuchiya_JMathPhys_39}. The determinants as well as their prefactors can be evaluated in the scaling limit of the XXZ spin chain to the Lieb-Liniger Bose gas. Within this limit a $q$-raised N\'eel state that contains finitely many down spins corresponds to the ground state of finitely many free bosons. This allows for a rigorous proof of the overlap formula of~\cite{LLpaper} for Lieb-Liniger Bethe states and a Bose-Einstein condensate (BEC) state with an arbitrary even number of particles. 
\end{abstract}

\section{Introduction}
Recently, there has been huge progress in understanding out-of-equilibrium dynamics in isolated many-body quantum systems, both theoretically and experimentally. Different experimental setups, in particular experiments on ultra cold atoms, can be found in the review article \cite{review} and references therein. 

Theoretically, one-dimensional integrable models play an essential role in under\-standing relaxation processes in many-body quantum systems. Many of them show strong correlations and there are (infinitely many) non-trivial conservation laws that strongly constrain the dynamics. Despite these constraints and although the time evolution is unitary, it is believed that local observables generically approach stationary values. The underlying assumption is that the system relaxes to a so-called generalized Gibbs ensemble \cite{2008_Rigol}, which depends upon as many parameters as the number of conservation laws. Calculating these parameters (even when the GGE is truncated) is, in general, a difficult problem \cite{2012_Mossel_JPA_45, 2013_Pozsgay}. In case of the one-dimensional spin-1/2 XXZ model one can avoid their explicit calculation by working with the so-called generating function \cite{Fagotti_1311.5216, 2013_Fagotti}, and using the quantum transfer matrix technique \cite{1993_Klumper, 2002_Klumper} and the related method of calculating short-distance correlation functions \cite{2007_Boos, 2008_Boos, 2010_Trippe}.

However, the recently-proposed quench action approach \cite{2013_Caux_PRL_110} does not utilize an underlying assumption for the steady state ensemble and hence circumvents those difficulties. Within this method the steady state can be exactly calculated in the thermodynamic limit. It determines stationary expectation values of operators as well as their full time dependencies. One of the main requirements for this approach is the knowledge of the large size scaling of overlaps of the initial state with energy eigenstates. 

For many integrable models the calculation of scalar products between different Bethe states \cite{1982_Korepin, 1984_Izergin, 1987_Izergin, 1990_Slavnov_TMP_79_82, 2007_Kitanine} is possible due to underlying algebraic structures (algebraic Bethe ansatz \cite{1979_Faddeev}, see e.g.~the textbook \cite{KorepinBOOK}). However, up to recent times very little was known about scalar products of eigenstates of different Hamiltonians. In two cases, namely the Lieb-Liniger Bose gas \cite{1963_Lieb_PR_130_1} and the one-dimensional spin-1/2 XXZ model \cite{1928_Heisi}, analytic expressions for such scalar products that are treatable in the thermodynamic limit were recently discovered \cite{XXZpaper, LLpaper}. They are all given by determinants of `Gaudin-like' form \cite{1981_Gaudin_PRD_23}. Hence, their scaling with large system size can be extracted, allowing then to study interaction quench problems following the approach of~\cite{LLpaper, 2013_Caux_PRL_110, XXZpaper2} as well as some thermodynamic equilibrium properties of spin chains as in~\cite{2012_Kozlowski_JSTAT_P05021}.

More specifically, in~\cite{LLpaper} the authors present a formula for overlaps of Lieb-Liniger Bethe states with the state of spatially uniformly distributed bosons (BEC state), the ground state of the non-interacting Bose gas. They checked their result analytically up to eight particles. In~\cite{XXZpaper} the same authors show that the overlap of the N\'eel state, the ground state of the antiferromagnetic Ising model, with XXZ Bethe states can be expressed similarly. One of the aims of the present paper is to give a rigorous proof of the Lieb-Liniger overlap formula for an arbitrary number of particles by using the proven results for XXZ overlaps of~\cite{XXZpaper}. 

In~\cite{XXZpaper} only the N\'eel state as an initial state is considered, which lies in the zero-magnetization sector of the XXZ spin chain. Here we shall present a formula for overlaps of Bethe states with so-called $q$-raised N\'eel states which lie in non-zero magnetization sectors. We consider overlaps with different initial states, namely the $q$-raised dimer and $q$-dimer states, as well.

The paper is organized as follows. In chapter \ref{sec:XXZ_basics} we define the main objects of the algebraic Bethe ansatz of the XXZ model and we present the most important formulas that are needed in following chapters. We introduce different initial states for which we can express the overlaps with XXZ Bethe states by a determinant of Gaudin type. We further discuss the special scaling limit of the XXZ spin chain which leads to the Lieb-Liniger Bose gas. In chapter \ref{sec:overlaps} we present the overlap formulas for those initial states. We show that one of these determinant expressions can be evaluated in the scaling limit to Lieb-Liniger which eventually proves the recently-proposed Lieb-Liniger overlap formula of~\cite{LLpaper}.

\section{Algebraic structures of the XXZ model and scaling limit to Lieb-Liniger}\label{sec:XXZ_basics}
The Hamiltonian of the one-dimensional spin-1/2 XXZ model is given by 
\begin{equation}\label{eq:Hamiltonian_XXZ}
	H = \sum_{j=1}^{N}\left(\sigma_{j}^{x}\sigma_{j+1}^{x}+\sigma_{j}^{y}\sigma_{j+1}^{y}+\Delta ( \sigma_{j}^{z}\sigma_{j+1}^{z}-1)\right)\epc
\end{equation}
where periodic boundary conditions $\sigma_{N+1}^{\alpha}=\sigma_{1}^{\alpha}$, $\alpha = x,y,z$, are supposed. We parametrize the anisotropy parameter of the model by $\Delta=\cosh(\eta)=(q+q^{-1})/2$. 

The XXZ model is integrable. It corresponds to a two-dimensional classical 6-vertex model \cite{BaxterBOOK}, which means that it is solvable using Bethe ansatz techniques \cite{1931_Bethe}, especially the algebraic Bethe ansatz \cite{1979_Faddeev, KorepinBOOK}.

One of the basic ideas of the algebraic Bethe ansatz is that the Hamiltonian of the model under consideration can be constructed as a member of an infinite series of conserved quantities which can be obtained from a family of commuting matrices. The transfer matrix $t$ depends on a spectral parameter $\lambda$ and is defined as the trace of the so-called monodromy matrix. The commutativity $[t(\lambda),t(\mu)]=0$ is related to an underlying algebraic structure of the monodromy matrix, the Yang-Baxter algebra, which is a set of quadratic relations defined by the so-called $R$-matrix. The latter can be interpreted as a vertex operator of a classical vertex model, which is, in case of the spin-1/2 XXZ chain, the $R$-matrix of the 6-vertex model \cite{BaxterBOOK}.

\subsection{Algebraic Bethe Ansatz for XXZ}
The integrable structure of the spin-1/2 XXZ chain is related to the Yang-Baxter algebra that is defined as the free associative algebra of generators $T^{\alpha}_{\beta}(\lambda)$, $\alpha,\beta=1,\ldots,d$, modulo the quadratic relations (see e.g.~the textbooks \cite{KorepinBOOK, HubbardBOOK})
\begin{equation}\label{eq:YBA}
	\check{R}(\lambda-\mu)\left(T(\lambda)\otimes T(\mu)\right) = \left(T(\mu) \otimes T(\lambda)\right)\check{R}(\lambda-\mu)\epp
\end{equation}
The $d\times d$ matrix $T(\lambda)$ is called monodromy matrix and has the generators of the Yang-Baxter algebra as entries. $\lambda$ is the spectral parameter. The R-matrix $\check{R}(\lambda)$ is a solution of the Yang-Baxter equation (in braid form) \cite{BaxterBOOK}
\begin{equation}\label{eq:YBE}
	\left(\check{R}(\lambda)\otimes\mathds{1}\right) \left(\mathds{1}\otimes \check{R}(\lambda+\mu)\right)\left(\check{R}(\mu)\otimes\mathds{1}\right) = \left(\mathds{1}\otimes \check{R}(\mu)\right)\left(\check{R}(\lambda+\mu)\otimes\mathds{1}\right)\left(\mathds{1}\otimes \check{R}(\lambda)\right)
\end{equation}
with the unity matrix $\mathds{1}$. 

In case of the XXZ model $d=2$ and the $R$-matrix is given by
\begin{equation}\label{eq:R_matrix}
	\check{R}(\lambda)=\frac{1}{\sinh(\lambda+\eta)}\left(\begin{array}{cccc}
		\sinh(\lambda+\eta) & 0 & 0 & 0\\
		0 & \sinh(\eta) & \sinh(\lambda) & 0\\
		0 & \sinh(\lambda) & \sinh(\eta) & 0\\
		0 & 0 & 0 & \sinh(\lambda+\eta)		
	\end{array}\right)\epc
\end{equation} 
which is the $R$-matrix of the 6-vertex model. The complex parameter $\eta$ is determined by the anisotropy parameter $\Delta = \cosh{\eta}$. For real $\eta \neq 0$ we have $\Delta> 1$ and we are in the antiferromagnetic gapped regime. For purely imaginary $\eta$ we have $-1\leq \Delta \leq 1$ and we are in the gapless regime. 

One can construct an explicit representation of the Yang-Baxter algebra \eqref{eq:YBA} using the explicit form of the $R$-matrix \eqref{eq:R_matrix}. Using the permutation operator $P$ and the Pauli matrices $\sigma^{\alpha}$, $\alpha=z,+,-$, the $R$-matrix can be written as 
\begin{equation}\label{eq:def_R_matrix}
	R(\lambda)=P\check{R	}(\lambda-\eta/2) = \frac{\sinh\left(\lambda+\frac{\eta}{2}\sigma^{z}\otimes\sigma^{z}\right)}{\sinh(\lambda+\eta/2)} +\frac{\sinh(\eta)\left(\sigma^{+}\otimes\sigma^{-} +\sigma^{-}\otimes\sigma^{+}\right)}{\sinh(\lambda+\eta/2)}\epp
\end{equation}
We introduce an auxiliary space $\mathbb{C}^2$ and index it with the letter $a$. We label the local quantum spaces with indices $n=1,\ldots,N$. The Lax operator on lattice site $n$ is defined as a $2\times 2$ matrix in the auxiliary space
\begin{equation}\label{eq:def_Lax}
	L_n(\lambda) = R_{an}(\lambda) = \frac{1}{\sinh(\lambda+\eta/2)}\left(\begin{array}{cc}
\sinh\big(\lambda+\frac{\eta}{2}\sigma_n^z\big) & \sinh(\eta)\sigma_n^- \\[1ex]
\sinh(\eta)\sigma_n^+ & \sinh\big(\lambda-\frac{\eta}{2}\sigma_n^z \big)	
	\end{array}\right)\epp
\end{equation}
The monodromy matrix is the product (in auxiliary space) of $N$ Lax operators \cite{KorepinBOOK}, 
\begin{equation}\label{eq:def_monodromy}
	T(\lambda)=\prod_{n=1}^N L_n(\lambda) = L_1(\lambda)\ldots L_N(\lambda)=: \left(\!\!\begin{array}{c@{\hspace{1.3ex}}c}
	A(\lambda) & B(\lambda) \\[0.7ex] C(\lambda) & D(\lambda)
	\end{array}\!\!\right)\epp
\end{equation}
It is a $2\times 2$ matrix with entries that are operators in the Hilbert space $(\mathbb{C}^2)^{\otimes N}$ of the XXZ spin chain, the $N$-fold tensor product of local spin-1/2 representation spaces $\mathbb{C}^2$. Using definitions \eqref{eq:def_R_matrix} and \eqref{eq:def_Lax} and the Yang-Baxter equation \eqref{eq:YBE} it is obvious that each Lax operator $L_n(\lambda)$, $n=1,\ldots,N$, is a representation of the Yang-Baxter algebra \eqref{eq:YBA}. Hence, the monodromy matrix \eqref{eq:def_monodromy} as a product of Lax operators acting on different lattice sites is a representation as well. 

The transfer matrix is defined as the trace over the auxiliary space of the monodromy matrix,
\begin{equation}\label{eq:transfer_matrix}	t(\lambda)=\text{tr}_a\big(T(\lambda)\big)=A(\lambda)+D(\lambda)\epp
\end{equation} 
Multiplying equation \eqref{eq:YBA} with the inverse of $\check{R}(\lambda-\mu)$ from the right and taking the trace on both sides we easily find that the transfer matrices build a commutative family,
\begin{equation}\label{eq:transfer_matrix_commutative_family}	t(\lambda)=\text{tr}\left(T(\lambda)\right) \quad\Rightarrow\quad [t(\lambda),t(\mu)]=0\epp
\end{equation}
From this commutativity one can easily see that the coefficients $J_m=\left.\frac{\partial^m}{\partial\lambda^m}\ln(t(\lambda))\right|_{\lambda=\eta/2}$ in an expansion of $\ln(t(\lambda))$ around $\lambda=\eta/2$ commute with each other. They are called the conserved currents of the XXZ spin chain and they form a commutative subalgebra of the Yang-Baxter algebra. Together with the explicit expression \eqref{eq:def_Lax} one finds that the Hamiltonian \eqref{eq:Hamiltonian_XXZ} is given by $J_1$,
\begin{equation}\label{eq:Hamiltonian_XXZ2}
	H = 2\sinh(\eta)J_1 = 2\sinh(\eta)\left.\frac{\partial}{\partial\lambda}\ln\left(t(\lambda)\right)\right|_{\lambda=\eta/2}\epp
\end{equation}

In order to construct eigenstates of the Hamiltonian (and all other conserved currents $J_m$) we need a pseudo vacuum $|0\rangle$ onto which the monodromy matrix acts triangularly, {\it i.e.}~$C(\lambda)|0\rangle = 0$. For this one can simply choose the fully-polarized state $|0\rangle=|\!\uparrow\ldots\uparrow\rangle=|\!\uparrow\rangle^{\otimes N}$. A Bethe state $|\{\lambda_j\}_{j=1}^M\rangle$ is defined as a product of $B$-operators from definition \eqref{eq:def_monodromy}, with (arbitrary) spectral parameters $\{\lambda_j\}_{j=1}^M$ that acts onto the pseudo vacuum,
\begin{equation}\label{eq:Bethe_state}
	|\{\lambda_j\}_{j=1}^{M}\rangle = \left[\prod\limits_{j=1}^M B(\lambda_j)\right]|0\rangle\epp
\end{equation} 
It is an eigenstate of the transfer matrix \eqref{eq:transfer_matrix}, and thus of the Hamiltonian \eqref{eq:Hamiltonian_XXZ}, if the parameters $\lambda_j$, $j=1,\ldots,M$, fulfill the Bethe equations
\begin{equation}\label{eq:BAE}
	\left(\frac{\sinh(\lambda_j+\eta/2)}{\sinh(\lambda_j-\eta/2)}\right)^N=-\prod_{k=1}^M\frac{\sinh(\lambda_j-\lambda_k+\eta)}{\sinh(\lambda_j-\lambda_k-\eta)}\epc \qquad j=1,\ldots,M \epp
\end{equation}
A solution $\{\lambda_j\}_{j=1}^M$ to these coupled algebraic equations with $\lambda_j \neq \lambda_k$ for all $j,k$ is called a set of Bethe roots. According to the conventions used in~\cite{XXZpaper} we shall call Bethe states `on-shell' if $\{\lambda_j\}_{j=1}^M$ is a set of Bethe roots, and `off-shell' otherwise. The eigenvalues of the transfer matrix $t$ and of the Hamiltonian $H$ are respectively given by
\begin{subequations}
\begin{align}
	\tau(\lambda) &= \prod_{k=1}^M\frac{\sinh(\lambda-\lambda_k-\eta)}{\sinh(\lambda-\lambda_k)} +  \left[\frac{\sinh\left(\lambda-\eta/2\right)}{\sinh\left(\lambda+\eta/2\right)}\right]^N\prod_{k=1}^M\frac{\sinh(\lambda-\lambda_k+\eta)}{\sinh(\lambda-\lambda_k)}\epc\\
	E &= 2\sinh(\eta)\left.\frac{\partial}{\partial\lambda}\ln\left(\tau(\lambda)\right)\right|_{\lambda=\eta/2} = \sum_{k=1}^M \frac{2\sinh^2(\eta)}{\sinh(\lambda_k+\eta/2)\sinh(\lambda_k-\eta/2)}\epp
\end{align}
\end{subequations}

A state of the form \eqref{eq:Bethe_state} is also an eigenstate of the magnetization $S^z = \sum_{n=1}^N \sigma_n^z/2$ with eigenvalue $N/2-M$. In the following we will call the space spanned by Bethe states with a fixed number $M$ of spectral parameters the sector of fixed magnetization $S^z=N/2-M$. Furthermore, a Bethe state is called parity invariant if the set of spectral parameters fulfills the symmetry $\{\lambda_j\}_{j=1}^M= \{-\lambda_j\}_{j=1}^M$. 

The norm of an on-shell Bethe state is given by 
\begin{subequations}\label{eq:norm_Bethe_state}
\begin{align}
\|\{\lambda_j\}_{j=1}^M\| &= \sqrt{\langle \{\lambda_j\}_{j=1}^M|  \{\lambda_j\}_{j=1}^M \rangle}\epc \\ 
\label{eq:norm_Bethe_state_b}
	\langle \{\lambda_j\}_{j=1}^M|  \{\lambda_j\}_{j=1}^M \rangle &= \sinh^M(\eta) \prod_{\substack{j,k=1\\j\neq k}}^M \frac{\sinh(\lambda_j - \lambda_k + \eta)}{\sinh(\lambda_j - \lambda_k)} \det{}_{\!M} (G_{jk}) \epc\\
	G_{jk} &= \label{eq:Gaudin_matrix}  \delta_{jk}\left(NK_{\eta/2}(\lambda_j)-\sum_{l=1}^{M}K_\eta(\lambda_j-\lambda_l)\right) + K_\eta(\lambda_j-\lambda_k)\epc
\end{align}
\end{subequations}
where $K_\eta(\lambda)=\frac{\sinh(2\eta)}{\sinh(\lambda+\eta)\sinh(\lambda-\eta)}$ is the derivative of $\theta(\lambda)=i\ln\big[\frac{\sinh(\lambda+\eta)}{\sinh(\lambda-\eta)}\big]$, the scattering matrix of the XXZ model. The norm formula was first suggested by Gaudin in~\cite{1981_Gaudin_PRD_23} and then rigorously proven by Korepin in~\cite{1982_Korepin}.

\subsection{Connection to the fundamental representation of $U_q(sl_2)$}
Let us consider $B$- and $C$-operators in the limit of an infinitely large spectral parameter. We write the monodromy matrix \eqref{eq:def_monodromy} in the following way ($q=e^\eta$, $s_n^z= \sigma_n^z/2$, $s_n^\pm = \sigma_n^\pm$):
\begin{align}
	\left(\!\!\begin{array}{c@{\hspace{1ex}}c}
	A(\lambda) & B(\lambda) \\[0.7ex] C(\lambda) & D(\lambda)
	\end{array}\!\!\right) 
	&= \frac{1}{\sinh^N(\lambda+\eta/2)}\prod_{n=1}^N \left(\!\begin{array}{c@{\hspace{0ex}}c}\sinh\big(\lambda+ \eta s_n^z\big) & \sinh(\eta)s_n^- \\[1.0ex] \sinh(\eta)s_n^+ & \sinh\big(\lambda- \eta s_n^z \big)	\end{array}\!\right)\notag\\[2ex]
	&\sim q^{\mp N/2}\prod_{n=1}^N \left[\left(\!\!\begin{array}{c@{\hspace{0.5ex}}c} q^{\pm s_n^z} & 0\\[0.5ex] 0 & q^{\mp s_n^z} \end{array}\!\!\right) 
	\pm 2e^{\mp\lambda}\sinh(\eta)\left(\!\begin{array}{c@{\hspace{0.5ex}}c} 0 & s_n^- \\[0.5ex] s_n^+ & 0\end{array}\!\right) \right]\epc
\end{align}
where the two signs indicate the different behavior for $\lambda\to\pm\infty$. Therefore, we get   
\begin{subequations}\label{eq:Sq_operators}
\begin{align}
	S_q^- &= 
	\lim_{\lambda\to+\infty} \left(\frac{q^{+N/2}\sinh(\lambda)B(\lambda)}{\sinh(\eta)} \right) = \sum_{n=1}^N \left[\prod_{j=1}^{n-1} q^{+s_j^z}\right] s_n^- \left[\prod_{j=n+1}^N q^{-s_j^z}\right]\epc \\
	\tilde{S}_q^- &= 
	\lim_{\lambda\to-\infty} \left( \frac{q^{-N/2}\sinh(\lambda)B(\lambda)}{\sinh(\eta)} \right) = \sum_{n=1}^N \left[\prod_{j=1}^{n-1} q^{-s_j^z}\right] s_n^- \left[\prod_{j=n+1}^N q^{+s_j^z}\right]\epc\\
	S_q^+ &= 
	\lim_{\lambda\to-\infty} \left( \frac{q^{-N/2}\sinh(\lambda)C(\lambda)}{\sinh(\eta)} \right) = \sum_{n=1}^N \left[\prod_{j=1}^{n-1} q^{+s_j^z}\right] s_n^+ \left[\prod_{j=n+1}^N q^{-s_j^z}\right]\epc\\
	\tilde{S}_q^+ &= 
	\lim_{\lambda\to+\infty} \left( \frac{q^{+N/2}\sinh(\lambda)C(\lambda)}{\sinh(\eta)} \right) = \sum_{n=1}^N \left[\prod_{j=1}^{n-1} q^{-s_j^z}\right] s_n^+ \left[\prod_{j=n+1}^N q^{+s_j^z}\right]\epc
\end{align}
\end{subequations}
which are $U_q(sl_2)$ symmetry operators \cite{1990_Pasquier}. We will use the raising and lowering operators $S_q^{\pm}$ and $\tilde{S}_q^{\pm}$ in the next section \ref{sec:initial_states} to create $q$-raised N\'eel and dimer states. 

The operators $q^{2s^z}$ and $s^\pm$ satisfy the relations $q^{2s^z}s^\pm = q^{\pm 2}s^\pm q^{2s^z}$ and $s^+s^--s^-s^+ = (q^{2s^z}-q^{-2s^z})/(q-q^{-1})$, and they form the so-called fundamental representation of $U_q(sl_2)$. Since $U_q(sl_2)$ has the structure of a Hopf algebra \cite{1985_Drinfeld, 1985_Jimbo} we can use the two co-multiplications, defined by
\begin{align*}
	\Delta(q^{2s^z}) &= q^{2s^z} \otimes q^{2s^z}\epc  & \Delta(s^+) &= q^{2s^z}\otimes s^+ + s^+\otimes\mathds{1}\epc  & \Delta(s^-) &= \mathds{1}\otimes s^- + s^-\otimes q^{-2s^z}\epc\\[1ex]
\tilde\Delta(q^{2s^z}) &= q^{2s^z} \otimes q^{2s^z}\epc  & \tilde\Delta(s^+) &=  \mathds{1}\otimes s^+ + s^+\otimes q^{2s^z}\epc & \tilde\Delta(s^-) &= q^{-2s^z}\otimes s^- + s^-\otimes \mathds{1}\epc
\end{align*}
to rewrite the operators $S_q^{\pm}$, $\tilde{S}_q^{\pm}$ in the following way 
\begin{subequations}\label{eq:S_operators}
\begin{align}
S_q^+ q^{1/2+\sum_{j=1}^Ns_j^z} 
&= \sum_{n=1}^N (q^{2s^z})^{\otimes n-1}\otimes s^+ \otimes \mathds{1}^{\otimes N-n} = \Delta^{N-1}(s^+)\epc\\
q^{-1/2-\sum_{j=1}^Ns_j^z}S_q^- 
&= \sum_{n=1}^N \mathds{1}^{\otimes n-1}\otimes s^- \otimes (q^{-2s^z})^{\otimes N-n} = \Delta^{N-1}(s^-)\epc\\ 
 \tilde{S}_q^+ q^{1/2+\sum_{j=1}^Ns_j^z} 
 &= \sum_{n=1}^N \mathds{1}^{\otimes n-1}\otimes s^+ \otimes (q^{2s^z})^{\otimes N-n} = \tilde\Delta^{N-1}(s^+)\epc\\
  q^{-1/2-\sum_{j=1}^Ns_j^z} \tilde{S}_q^- 
  &= \sum_{n=1}^N (q^{-2s^z})^{\otimes n-1}\otimes s^- \otimes \mathds{1}^{\otimes N-n} = \tilde\Delta^{N-1}(s^-)\epp
 \end{align}
\end{subequations}
We will need this representation of $q$-raising and $q$-lowering operators to investigate the limit $q\to-1$ in section~\ref{sec:scaling_limit}. There are in principle problems with periodic boundary conditions, which is discussed in~\cite{1990_Pasquier}, but for our purposes, especially in the limits $N\to\infty$ and $q\to-1$, they can be ignored.

\subsection{Different initial states}\label{sec:initial_states}
We are interested in overlaps of special initial states $\left|\Psi\right\rangle$ with parity-invariant Bethe states. Note that for some of the initial states also non parity-invariant Bethe states have non-zero overlap which is important for non-equilibrium dynamics. For convenience we choose in the following $N$ divisible by four and $M$ even, and denote parity invariant Bethe states by $|\{\pm\lambda_j\}_{j=1}^{m}\rangle$, $m=M/2$. We want to calculate overlaps $\langle \Psi | \{\pm\lambda_j\}_{j=1}^{m}\rangle$. Some states for which a Gaudin-like determinant expression exists (see section~\ref{sec:overlaps}) are \cite{Pozsgay_1309.4593}
\begin{subequations}\label{eq:initial_states}
\begin{itemize}
	\item the N{\'e}el and the anti-N\'eel state 
	\begin{equation}\label{eq:Neel}
		\left|\Psi_N\right\rangle = \left|\uparrow\downarrow \uparrow\downarrow\ldots\right\rangle\epc \qquad \left|\Psi_{AN}\right\rangle=\left|\downarrow\uparrow\downarrow\uparrow \ldots \right\rangle\epc
\end{equation}	
and especially its symmetric combination $\left|\Psi_0\right\rangle = \frac{1}{\sqrt{2}}(\left|\Psi_N\right\rangle + \left|\Psi_{AN}\right\rangle)$, which we call the zero-momentum N{\'e}el state,
\item the dimer state 
	\begin{equation}\label{qe:dimer}
		\left|\Psi_D\right\rangle = \bigotimes_{j=1}^{N/2}\frac{\left|\uparrow\downarrow\right\rangle-\left|\downarrow\uparrow\right\rangle}{\sqrt{2}}\epc
	\end{equation}	
\item and the q-dimer state
\begin{equation}\label{eq:qdimer}
		\left|\Psi_{qD}\right\rangle = \bigotimes_{j=1}^{N/2}\frac{q^{1/2}\left|\uparrow\downarrow\right\rangle-q^{-1/2}\left|\downarrow\uparrow\right\rangle}{\sqrt{|q|+|q|^{-1}}}\epc
	\end{equation}	
	where here and in the following the value of $q$ is fixed by the anisotropy parameter of the Hamiltonian \eqref{eq:Hamiltonian_XXZ}, $\Delta = \cosh(\eta) = (q+q^{-1})/2$.
\end{itemize} 
\end{subequations}
They all lie in the sector of zero magnetization, $S^z=N-M/2=0$, and they only have non-vanishing overlaps with Bethe states $|\{\pm\lambda_j\}_{j=1}^{m}\rangle$ if $m=M/2=N/4$. The corresponding (unnormalized) $2n$-fold $q$-raised states are
\begin{subequations}\label{eq:q-raised_states}
\begin{itemize}
	\item the $q$-raised N\'eel state
\begin{equation}\label{eq:q-raised_Neel}
	|\Psi_N^{(n)}\rangle = \left(S_q^+\tilde{S}_q^+\right)^{n}\left|\Psi_N\right\rangle\epc
\end{equation} 
	\item the $q$-raised dimer state
\begin{equation}\label{eq:q-raised_dimer}
	|\Psi_{D}^{(n)}\rangle = \left(S_q^+\tilde{S}_q^+\right)^{n}\left|\Psi_D\right\rangle\epc
\end{equation} 
\item and the $q$-raised $q$-dimer state
\begin{equation}\label{eq:q-raised_qdimer}
	|\Psi_{qD}^{(n)}\rangle = \left(S_q^+\right)^{2n}\left|\Psi_{qD}\right\rangle\epp
\end{equation} 
\end{itemize}
\end{subequations}
Note that $\tilde{S}_q^\pm|\Psi_{qD}\rangle = 0$. All of these initial states have non-vanishing magnetization $S^z=2n$ and we necessarily have $m=M/2=N/4-n$. To calculate overlaps of these states with Bethe ket states we need the corresponding bra states. Since $\left(S_q^+\right)^\dagger= S_q^-$ and $\left(\tilde{S}_q^+\right)^\dagger= \tilde{S}_q^-$ commute, they can be simply obtained by acting with $\left(S_q^-\tilde{S}_q^-\right)^{n}$, $\left(S_q^-\right)^{2n}$ from the right on $\left\langle\Psi_{X}\right|$ for $X=N, D, qD$, respectively. 

\subsection{Scaling limit to the Lieb-Liniger model}\label{sec:scaling_limit}
The scaling limit of the spin-1/2 XXZ chain to the Lieb-Liniger Bose gas is given by \cite{GaudinBOOK, 1987_Golzer, 2007_Seel, Pozsgay_JStatMech_P11017}
\begin{equation}\label{eq:scaling_limit}
	\eta = i\pi - i\epsilon\epc \qquad N = cL/\epsilon^2\epc\qquad \lambda_j \to \epsilon\lambda_j/c\epc\qquad \epsilon\to 0\epp
\end{equation}
The Bethe equations \eqref{eq:BAE} for a finite number $M$ of rapidities become ($N$ even)
\begin{equation}
	e^{iL\lambda_j} = -\prod_{k=1}^M\frac{\lambda_j-\lambda_k+ic}{\lambda_j-\lambda_k-ic}\epc \qquad j=1,\ldots,M\epp
\end{equation}
These are the Bethe equations of the Lieb-Linger model \cite{1963_Lieb_PR_130_1}. 

Since $q=e^\eta\to-1$ in the limit \eqref{eq:scaling_limit} the operators $S_q^\pm$, $\tilde{S}_q^\pm$ become staggered $SU(2)$ symmetry operators (up to a trivial prefactor)
\begin{equation}
	S_{q}^\pm, \tilde{S}_{q}^\pm \to S_{\text{st}}^\pm=\sum_{n=1}^N(-1)^n s_n^\pm\epc
\end{equation}
which can be seen using the representation \eqref{eq:S_operators} for the $q$-raising operators. Since the operators $s_n^\pm$ act locally as spin raising and spin lowering operators and as they act non-trivially only on even or only on odd lattice sites, $S^\pm_{\text{st}}$ act on the N\'eel state \eqref{eq:Neel} as usual global $SU(2)$ spin raising and lowering operators $S^\pm$. We eventually obtain for the ($N/2-2m$)-fold $q$-raised N\'eel state \eqref{eq:q-raised_Neel}
\begin{equation}
	\left\langle \Psi_N \right|(S_q^{-}\tilde{S}_q^{-})^{N/4-m} = \left\langle \Psi_N \right|(S^{-})^{N/2-2m}\epp
\end{equation}
This is a state with $2m$ uniformly-distributed down spins. In the scaling limit to Lieb-Liniger it corresponds to the state of $N_{LL}=2m$ spatially uniformly-distributed bosons, the so-called BEC state of~\cite{LLpaper}.

\section{Overlaps for $q$-raised states}\label{sec:overlaps}
In order to obtain an expression for the overlap of a $q$-raised state \eqref{eq:q-raised_states} with a normalized parity-invariant XXZ on-shell Bethe state in the non-zero magnetization sector we start in section~\ref{sec:old_overlap} with the overlap of a zero magnetization state \eqref{eq:initial_states} with an unnormalized parity-invariant off-shell state $|\{\pm\lambda_j\}_{j=1}^{N/4} \rangle$ as in~\cite{XXZpaper}. The calculation of overlaps of $q$-raised states can be reduced to the calculation of overlaps in the zero-magnetization sector where, according to equation~\eqref{eq:Sq_operators}, some of the spectral parameters of the Bethe state, $\{\mu_j\}_{j=1}^{2n}$, $2n=N/2-2m$, are sent to infinity,
\begin{subequations}
\begin{align}\label{eq:OL_q_to_normal}
	\langle \Psi_{N,D}^{(n)} | \{\pm\lambda_j\}_{j=1}^{m}\rangle &= \lim\nolimits_{\{\mu_j\to\infty\}_{j=1}^{n}} (-1)^n\prod_{j=1}^n\frac{\sinh^2(\mu_j)}{\sinh^2(\eta)}\langle \Psi_{N,D} | \{\pm\lambda_j\}_{j=1}^{m}\cup\{\pm\mu_j\}_{j=1}^{n}\rangle\epc\\
	\langle \Psi_{qD}^{(n)} |\{\pm\lambda_j\}_{j=1}^{m}\rangle &= \lim\nolimits_{\{\mu_j\to\infty\}_{j=1}^{2n}}q^{nN}\prod_{j=1}^{2n}\frac{\sinh(\mu_j)}{\sinh(\eta)}\langle \Psi_{qD} | \{\pm\lambda_j\}_{j=1}^{m}\cup\{\mu_j\}_{j=1}^{2n}\rangle\epp
\end{align} 
\end{subequations}
The minus sign $(-1)^n$ in the first equation comes from the fact that the Bethe state is parity-invariant and that we always send two parameters $\pm \mu_j$ to $\pm\infty$ at the same time.

\subsection{Connection between different initial states}
In~\cite{Pozsgay_1309.4593} it is shown that the overlaps of the different initial states \eqref{eq:initial_states} with an XXZ Bethe state (not necessarily parity invariant) are related to each other, 
\begin{multline}\label{eq:OL_rel_normal}
	\langle \Psi_N|\{\lambda_j\}_{j=1}^{N/2}\rangle\prod_{j=1}^{N/2}\frac{\sinh(\eta)/\sqrt{2}}{\sinh(\eta/2+\lambda_j)} = \langle \Psi_D|\{\lambda_j\}_{j=1}^{N/2}\rangle \prod_{j=1}^{N/2}\frac{\cosh(\frac{\eta}{2})}{\cosh(\lambda_j)} \\ = \langle \Psi_{qD}|\{\lambda_j\}_{j=1}^{N/2}\rangle \prod_{j=1}^{N/2}\frac{\sqrt{\cosh(\eta)}}{\exp(\lambda_j)}\epp
\end{multline}
The last equation is only true for $\Delta>1$. For $\Delta <1$ the square root $\sqrt{\cosh(\eta)}$ disappears. Similar relations are true for the corresponding $q$-raised states \eqref{eq:q-raised_states}. In this case we send pairs of parameters to infinity: $\pm\mu_j\to\pm\infty$, $j=1,\ldots,n$, for the N\'eel and the dimer state, and $\mu_{j}\to\infty$, $j=1,\ldots,2n$ for the $q$-dimer state. The rest of the parameters belong to the parity-invariant Bethe state denoted by $|\lambda_\pm\rangle = |\{\pm\lambda_j\}_{j=1}^{m}\rangle$. The divergent factors cancel and we obtain the relations ($m+n=N/4$)
\begin{equation}\label{eq:OL_rel_qdimer}
	\frac{(-2)^{N/4}\langle \Psi_N^{(n)}|\lambda_{\pm}\rangle}{[\sinh^2(\eta)]^{-n}}\prod_{j=1}^{m}E(\lambda_j) = \frac{\langle \Psi_D^{(n)}|\lambda_\pm\rangle}{[\cosh^2(\frac{\eta}{2})]^{-n}} \prod_{j=1}^{m}\frac{\cosh^2(\frac{\eta}{2})}{\cosh^2(\lambda_j)} = \frac{\langle \Psi_{qD}^{(n)}|\lambda_\pm\rangle}{[\cosh(\eta)]^{-n-m}}
\end{equation}
with $E(\lambda) = \frac{\sinh^2(\eta)}{\sinh(\lambda+\eta/2)
\sinh(\lambda-\eta/2)}$. 
Furthermore, in the scaling limit \eqref{eq:scaling_limit}, all factors in front of the overlaps become independent of the rapidities $\{\lambda_j\}_{j=1}^m$, and the relation between the overlaps of the $q$-raised N\'eel state and of the $q$-raised dimer state just reads
\begin{equation}
	\langle \Psi_D^{(n)}|\lambda_{\pm}\rangle = 2^{N/4}\langle \Psi_N^{(n)}|\lambda_\pm\rangle\epp
\end{equation}
Due to this relation we only need to consider in the following $q$-raised N\'eel states.

\subsection{Overlap of the N{\'e}el state with an off-shell Bethe state}\label{sec:old_overlap}
The overlap of the N\'eel state \eqref{eq:Neel} with an unnormalized parity-invariant XXZ off-shell state \eqref{eq:Bethe_state} is given by \cite{XXZpaper}
\begin{subequations}\label{eq:overlap_XXZ_offshell} 
\begin{equation}
	 \langle \Psi_N |\{\pm\lambda_j\}_{j=1}^{N/4}\rangle =  \gamma\det{}_{\!N/4}(G_{jk}^{+})\epc
\end{equation}
where the prefactor $\gamma$ and the matrix $G_{jk}^+$ read
\begin{align}\label{eq:gamma}
\gamma &= \left[\prod_{j=1}^{N/4}\frac{\sine{\lambda_j}{+\eta/2}\sine{\lambda_j}{-\eta/2}}{\sine{2\lambda_j}{0}^2}\right]\left[\prod_{\substack{j>k=1\\ \ \sigma=\pm}}^{N/4}\frac{\sine{\lambda_j+\sigma\lambda_k}{+\eta}\sine{\lambda_j+\sigma\lambda_k}{-\eta}}{\sine{\lambda_j+\sigma\lambda_k}{0}^2}\right]\\[3ex] 
G_{jk}^{+} &= \delta_{jk}\left(N\sine{0}{\eta}K_{\eta/2}(\lambda_j)-\sum_{l=1}^{N/4}\sine{0}{\eta}K_\eta^{+}(\lambda_j,\lambda_l)\right) + \sine{0}{\eta}K_\eta^{+}(\lambda_j,\lambda_k)\notag \\ \label{eq:Gaudin_plus}
&\qquad\quad + \delta_{jk}\frac{\sine{2\lambda_j}{+\eta}\,\mathfrak{A}_j+\sine{2\lambda_j}{-\eta}\,\bar{\mathfrak{A}}_j}{\sine{2\lambda_j}{0}} + (1-\delta_{jk})f_{jk}\epc \quad\qquad j,k=1,\ldots,N/4\\[3ex] \label{eq:f_jk}
f_{jk} 
&= \mathfrak{A}_k\left( \frac{\sine{2\lambda_j}{+\eta} \sine{0}{\eta}}{\sine{\lambda_j+\lambda_k}{0}\sine{\lambda_j-\lambda_k}{+\eta}} - \frac{\sine{2\lambda_j}{-\eta}\sine{0}{\eta}}{\sine{\lambda_j-\lambda_k}{0}\sine{\lambda_j+\lambda_k}{-\eta}} \right) + \mathfrak{A}_k\bar{\mathfrak{A}}_j \left(\frac{\sine{2\lambda_j}{-\eta}\sine{0}{\eta}}{\sine{\lambda_j-\lambda_k}{0}\sine{\lambda_j+\lambda_k}{-\eta}}\right) \notag\\
&\quad - \bar{\mathfrak{A}}_j\left(\frac{\sine{2\lambda_j}{-\eta}\sine{0}{\eta}}{\sine{\lambda_j-\lambda_k}{0}\sine{\lambda_j+\lambda_k}{-\eta}} + \frac{\sine{2\lambda_j}{-\eta}\sine{0}{\eta}}{\sine{\lambda_j+\lambda_k}{0}\sine{\lambda_j-\lambda_k}{-\eta}}\right)
\end{align}
with $K_\eta^{+}(\lambda,\mu)=K_\eta(\lambda+\mu)+K_\eta(\lambda-\mu)$ and $K_\eta(\lambda)=\frac{\sine{0}{2\eta}}{\sine{\lambda}{+\eta}\sine{\lambda}{-\eta}}$. We also introduced the shortcuts $\sine{\lambda}{\eta}=\sinh(\lambda+\eta)$ and
\begin{equation}\label{eq:func_a_tilde}
	\mathfrak{A}_j = 1 + \mathfrak{a}_j\epc\quad \bar{\mathfrak{A}}_j = 1 + \mathfrak{a}_j^{-1}\epc\quad \mathfrak{a}_j = \left[\prod_{\substack{k=1\\ \ \sigma=\pm}}^{N/4}\frac{\sine{\lambda_j-\sigma\lambda_k}{-\eta}}{\sine{\lambda_j-\sigma\lambda_k}{+\eta}}\right]\left(\frac{\sine{\lambda_j}{+\eta/2}}{\sine{\lambda_j}{-\eta/2}}\right)^{N}\epp
\end{equation}
\end{subequations}
Note that there is a difference of a factor $\sqrt{2}$ in $\gamma$ as compared to~\cite{XXZpaper} since here we consider the N\'eel state instead of the symmetric combination of N\'eel and anti-N\'eel. Here the parameters $\lambda_j$, $j=1,\ldots, N/4$, are arbitrary complex numbers.

\subsection{Determinant expression for the overlap of a q-raised N\'eel state}
First, we split the set of rapidities in formula \eqref{eq:overlap_XXZ_offshell} into two subsets labeled by $\{\pm\lambda_j\}_{j=1}^m$ and $\{\pm\mu_j\}_{j=1}^n$, $m+n=N/4$. To get the overlap of the $(N/2-2m)$-fold $q$-raised N\'eel state $\langle\Psi_N^{(N/4-m)}|$ with a parity-invariant Bethe state $|\{\pm\lambda_j\}_{j=1}^m\rangle$, we then have to take the limits $\mu_j\to\infty$ as in equation~\eqref{eq:OL_q_to_normal}. We shall do this step by step. 

We start with the $\mu$-dependent part of the factor $\gamma$ in equation~\eqref{eq:gamma}. Together with the normalization factor in equation~\eqref{eq:OL_q_to_normal} it becomes
\begin{equation}\label{eq:gamma_mu_to_infty}
	\gamma_\mu = (-1)^n\left[\prod_{j=1}^n\frac{\sine{\mu_j}{0}^2}{\sine{0}{\eta}^2}\right]\left[\prod_{j=1}^{n}\frac{\sine{\mu_j}{+\eta/2}\sine{\mu_j}{-\eta/2}}{\sine{2\mu_j}{0}^2}\right]\left[\prod_{\substack{j>k=1\\ \ \sigma=\pm}}^{n}\frac{\sine{\mu_j+\sigma\mu_k}{+\eta}\sine{\mu_j+\sigma\mu_k}{-\eta}}{\sine{\mu_j+\sigma\mu_k}{0}^2}\right]\epc
\end{equation} 
where we already neglected in the last product all factors containing one $\mu$- and one $\lambda$-parameter since they all become unity in the limit $\mu\to\infty$. We now send the $\mu$-parameters to infinity in such a way that all differences and sums of $\mu$'s are infinity. Then the third product becomes unity as well and we have $\gamma_\mu \to \gamma_\infty = (-1)^n4^{-n}\sine{0}{\eta}^{-2n}$. In total,
\begin{equation}\label{eq:new_gamma}
	\gamma = \gamma_\infty\hat{\gamma}=\frac{(-1)^n}{4^n\sine{0}{\eta}^{2n}} \left[\prod_{j=1}^{m}\frac{\sine{\lambda_j}{+\eta/2}\sine{\lambda_j}{-\eta/2}}{\sine{2\lambda_j}{0}^2}\right]\left[\prod_{\substack{j>k=1\\ \ \sigma=\pm}}^{m}\frac{\sine{\lambda_j+\sigma\lambda_k}{+\eta}\sine{\lambda_j+\sigma\lambda_k}{-\eta}}{\sine{\lambda_j+\sigma\lambda_k}{0}^2}\right]\epp
\end{equation}

The second step is to calculate the determinant of the matrix $G_{jk}^+$ in equation~\eqref{eq:overlap_XXZ_offshell} in the limits $\mu_j\to\infty$, $j=1,\ldots,n$. We immediately see that all $K^{+}$-terms vanish as long as one of the two arguments is one of the $\mu$-parameters. Furthermore, in the first $m$ rows and last $n$ columns, {\it i.e.}~$\lambda_j$ finite and $\lambda_k=\mu_{k-m}$, the terms $f_{jk}$ in equation~\eqref{eq:f_jk} vanish since the symbols $\mathfrak{A}_j$ are bounded and all factors inside the brackets vanish. Hence, the entire upper right $m\times n$ block of the matrix $G_{jk}^+$ is zero, and the determinant becomes decomposed into the product of two determinants. One is just the determinant of a reduced Gaudin-like matrix, 
\begin{align}
	\hat{G}_{jk}^{+} &= \delta_{jk}\left(N\sine{0}{\eta}K_{\eta/2}(\lambda_j)-\sum_{l=1}^{m}\sine{0}{\eta}K_\eta^{+}(\lambda_j,\lambda_l)\right) + \sine{0}{\eta}K_\eta^{+}(\lambda_j,\lambda_k)\notag \\ \label{eq:new_Gaudin_matrix}
&\qquad\quad + \delta_{jk}\frac{\sine{2\lambda_j}{+\eta}\,\mathfrak{A}_j+\sine{2\lambda_j}{-\eta}\,\bar{\mathfrak{A}}_j}{\sine{2\lambda_j}{0}} + (1-\delta_{jk})f_{jk}\epc \quad\qquad j,k=1,\ldots,m\epc
\end{align}
where $K_\eta$, $K_\eta^+$, $f_{jk}$, $\mathfrak{A}_j = 1 +\mathfrak{a}_j$, $\bar{\mathfrak{A}}_j = 1 +\mathfrak{a}_j^{-1}$ are defined as before (see equations~\eqref{eq:f_jk},~\eqref{eq:func_a_tilde}) and the symbols $\mathfrak{a}_j$, $j=1,\ldots,m$, reduce to
\begin{equation}\label{eq:new_func_a_tilde}
	\mathfrak{a}_j = \left[\prod_{\substack{k=1\\ \ \sigma=\pm}}^{m}\frac{\sine{\lambda_j-\sigma\lambda_k}{-\eta}}{\sine{\lambda_j-\sigma\lambda_k}{+\eta}}\right]\left(\frac{\sine{\lambda_j}{+\eta/2}}{\sine{\lambda_j}{-\eta/2}}\right)^{N}\epp
\end{equation}
The other determinant can be easily evaluated. Fixing a special order of limits $\mu_j\to\infty$, $j=1,\ldots,n$, in such a way that $\mu_k-\mu_j\to+\infty$ for $j>k$, the lower right $n\times n$ block of the matrix $G_{jk}^+$ becomes a triangular matrix and the determinant is just the product of all diagonal elements $D_j$. We thus have $\det{}_{\!M/2}(G_{jk}^+) = \det{}_{\!m}(\hat{G}_{jk}^+)\prod_{j=1}^n D_j$.

The next task is to calculate these diagonal elements. Using the previously-introduced order of limits, which we denote by  $\lim\nolimits_{\mu}$, we obtain 
\begin{equation}
	\mathfrak{a}_j=\lim\nolimits_\mu\left\{ \left[\prod_{\substack{k=1\\ \ \sigma=\pm}}^{n}\frac{\sine{\mu_j-\sigma\mu_k}{-\eta}}{\sine{\mu_j-\sigma\mu_k}{+\eta}}\right] 
	\left[\prod_{\substack{k=1\\ \ \sigma=\pm}}^{m}\frac{\sine{\mu_j-\sigma\lambda_k}{-\eta}}{\sine{\mu_j-\sigma\lambda_k}{+\eta}}\right]
	\left(\frac{\sine{\mu_j}{+\eta/2}}{\sine{\mu_j}{-\eta/2}}\right)^{N} \right\} = -e^{4\eta(j-1/2)}\epp
\end{equation}
Therefore, the diagonal elements can be written as [see the third term in equation~\eqref{eq:Gaudin_plus}], 
\begin{multline}\label{eq:new_factor}
	D_j = e^\eta \mathfrak{A}_j + e^{-\eta}\bar{\mathfrak{A}}_j = e^\eta (1-e^{4\eta(j-1/2)}) + e^{-\eta}(1-e^{-4\eta(j-1/2)})\\ = -4\sinh((2j-1)\eta)\sinh(2j\eta)\epp
\end{multline}
Together with the $n$-dependent part $\gamma_\infty$ of the $\gamma$-factor the product of all diagonal elements becomes
\begin{equation}
	\frac{(-1)^n}{4^n\sinh^{2n}(\eta)}\prod_{j=1}^n D_j = \prod_{j=1}^{n}\frac{\sinh((2j-1)\eta)\sinh(2j\eta)}{\sinh^2(\eta)} = \prod_{j=1}^{2n}\frac{q^j-q^{-j}}{q-q^{-1}} = [2n]_q!\epp
\end{equation}

As a final result we obtain the overlap of the normalized $q$-raised N\'eel state with normalized parity-invariant on-shell Bethe states. All together, using norm formula \eqref{eq:norm_Bethe_state} of an on-shell Bethe state, we have [$\hat{\gamma}$ and $\hat{G}_{jk}$ are defined in equations~\eqref{eq:new_gamma} and \eqref{eq:new_Gaudin_matrix}]
\begin{subequations}\label{eq:new_overlap}
\begin{align}\label{eq:new_overlap_a}
	 \frac{\langle \Psi_N^{(n)} |\{\pm\lambda_j\}_{j=1}^{m}\rangle}{\|\Psi_N^{(n)} \| \|\{\pm\lambda_j\}_{j=1}^m\|} &=  \frac{[2n]_q!}{\|\Psi_N^{(n)}\|} \frac{\hat{\gamma}\det{}_{\!m}(\hat{G}_{jk}^{+})}{\|\{\pm\lambda_j\}_{j=1}^m\|} \notag \\
&= \frac{[2n]_q!}{\|\Psi_N^{(n)}\|} \left[\prod_{j=1}^{m}\frac{\sqrt{\tanh(\lambda_j+\frac{\eta}{2}) \tanh(\lambda_j-\frac{\eta}{2})}}{2\sinh(2\lambda_j)}\right]\sqrt{ \frac{\det_{m}(\hat{G}_{jk}^{+})}{\det_{m}(\hat{G}_{jk}^{-})}}
\end{align}
where
\begin{equation}
\hat{G}_{jk}^\pm = \delta_{jk}\left(NK_{\eta/2}(\lambda_j)-\sum_{l=1}^{m}K_\eta^+(\lambda_j,\lambda_l)\right) + K_\eta^\pm(\lambda_j,\lambda_k)\epc \quad j,k=1,\ldots,m\epc
\end{equation}
\end{subequations}
and $K_\eta^\pm$, $K_\eta$ are defined as before. Here the parameters $\lambda_j$, $j=1,\ldots,m$, are Bethe roots but still, in general, complex numbers (string solutions). $\|\Psi_N^{(n)}\|$ is the norm of the $2n$-fold $q$-raised N\'eel state. We calculate this norm in the limit $q\to -1$ in section~\ref{sec:proof_LL}. 

We can use overlap formula \eqref{eq:new_overlap} for $q$-raised N{\'e}el states to prove the formula for overlaps of Lieb-Liniger Bethe states with the BEC state of one-dimensional free Bosons, which was recently discovered in~\cite{LLpaper}.

\subsection{Scaling to Lieb-Liniger and proof of the BEC Lieb-Liniger overlap formula}\label{sec:proof_LL}
In this section we prove the Lieb-Liniger overlap formula of~\cite{LLpaper} for an arbitrary even number of bosons. We have already seen at the end of chapter \ref{sec:scaling_limit} that, in the scaling limit of the XXZ spin chain to the Lieb-Liniger Bose gas, the $(N/2-N_{LL})$-fold $q$-raised N\'eel state scales to the BEC state of $N_{LL}$ bosons. We investigate in the following the identification of these states. The normalized BEC state is given by the $N\to\infty$ limit of the state with $N_{LL}$ uniformly distributed down spins,
\begin{equation}
	|BEC\rangle\ \widehat{=}\ \begin{pmatrix} N \\ N_{LL}\end{pmatrix}^{-1/2}\sum_{\{n_j\}_{j=1}^{N_{LL}}}\sigma^-_{n_1}\ldots\sigma_{n_{N_{LL}}}^-|\uparrow\ldots\uparrow\rangle\epp
\end{equation}
The sum is over all $N \choose N_{LL}$ subsets $\{n_j\}_{j=1}^{N_{LL}}$ of the first $N$ integers. The normalized $(N/2-N_{LL})$-fold $q$-raised N\'eel state reads
\begin{equation}
	\frac{ (S^+)^{N/2-N_{LL}}|\Psi_N\rangle}{\| (S^+)^{N/2-N_{LL}}|\Psi_N\rangle \| }  = \begin{pmatrix} N/2 \\ N_{LL}\end{pmatrix}^{-1/2}\sum_{\substack{\{n_j\}_{j=1}^{N_{LL}} \\ n_j \text{ even}}} \sigma_{n_1}^-\ldots\sigma_{n_{N_{LL}}}^-|\uparrow\ldots\uparrow\rangle\epp
\end{equation} 
Here the sum is over all $N/2 \choose N_{LL}$ subsets of even integers from $1$ to $N$ because in the $q$-raised N\'eel state the down spins sit only on even lattice sites. In the large $N$ limit the ratio of numbers of local spin basis states can be calculated by means of Stirling's formula (note that $N_{LL}$ is finite),
\begin{equation}
	\lim_{N\to\infty}\left[ \begin{pmatrix} N \\ N_{LL}\end{pmatrix} \left/ \begin{pmatrix} N/2 \\ N_{LL} \end{pmatrix}\right.\right] = \lim_{N\to\infty}\left[\frac{N!\:(\frac{N}{2}-N_{LL})!}{(\frac{N}{2})!(N-N_{LL})!}\right] = 2^{N_{LL}}\epp
\end{equation}

In the scaling limit we can identify the $q$-raised N\'eel state itself with the BEC state. In order to do this we have to multiply overlaps of this state with a factor $2^{N_{LL}}$ that takes account of the contribution of all `missing' states which also scale to the BEC state in the dilute limit. We further have to divide by a factor $\sqrt{2^{N_{LL}}}$ that corrects for the norm of the state. Both factors together therefore lead to a corrective factor of $2^{N_{LL}/2}$.

Between XXZ and Lieb-Liniger Bethe states there is a one-to-one correspondence \cite{1987_Golzer}. Furthermore, the Gaudin matrix of norm formula \eqref{eq:norm_Bethe_state} turns into the Gaudin matrix of the Lieb-Liniger norm \cite{KorepinBOOK}. Similarly, the modified Gaudin matrices $\hat{G}_{jk}^\pm$ turn into the corresponding Lieb-Liniger matrices. 

The norm of the initial state is given by $\|\Psi_N^{(n)}\| = (2n)! \sqrt{N/2 \choose 2n}= (2n)! \sqrt{N/2 \choose 2m}$. Using the scaling limit \eqref{eq:scaling_limit} we obtain for the prefactor in equation~\eqref{eq:new_overlap_a}, where we omit a factor $(-1)^n$ coming from the $q$-deformed factorial when $q\to-1$, and where we use the corrective factor $2^{N_{LL}/2}$, 
\begin{align}
	2^{N_{LL}/2}\frac{[2n]_q!}{\|\Psi_N^{(n)}\|} & \left[\prod_{j=1}^{m} 
	\frac{\sqrt{\tanh(\lambda_j+\frac{\eta}{2}) \tanh(\lambda_j-\frac{\eta}{2})}}{2\sinh(2\lambda_j)}\right] \notag\\
	&\to \frac{2^{N_{LL}/2}}{\sqrt{N/2 \choose 2m}} \left[\prod_{j=1}^{m}\frac{\sqrt{\coth(\epsilon\lambda_j/c-\frac{i\epsilon}{2}) \coth(\epsilon\lambda_j/c+\frac{i\epsilon}{2})}}{2\sinh(2\epsilon\lambda_j/c)}\right] \notag \\[1ex]
	&\to \frac{2^{N_{LL}/2}}{4^m\epsilon^{2m}}\sqrt{\frac{(N/2-2m)!}{(N/2)!}}\sqrt{(2m)!} \left[\prod_{j=1}^{m}\frac{1}{\frac{\lambda_j}{c}\sqrt{\frac{\lambda_j^2}{c^2}+\frac{1}{4}}}\right] \notag \\[2ex]
	&\to \frac{2^{N_{LL}/2}}{N^m\epsilon^{2m}}\frac{\sqrt{(2m)!}}{2^m} \left[\prod_{j=1}^{m}\frac{1}{\frac{\lambda_j}{c}\sqrt{\frac{\lambda_j^2}{c^2}+\frac{1}{4}}}\right] 
	\to \frac{(cL)^{-N_{LL}/2}\sqrt{N_{LL}!}}{\displaystyle \prod_{j=1}^{N_{LL}/2}\frac{\lambda_j}{c}\sqrt{\frac{\lambda_j^2}{c^2}+\frac{1}{4}}}\epp
\end{align}
In the second last step we used Stirling's formula and in the last step we plugged in $N=cL/\epsilon^2$ and $m=N_{LL}/2$. We eventually combine this with the determinants in the scaling limit to 
\begin{equation}\label{eq:Overlaps}
\frac{\langle  BEC | \{\pm\lambda_j\}_{j=1}^{N_{LL}/2} \rangle}{\|\{\pm\lambda_j\}_{j=1}^{N_{LL}/2} \|} = \frac{\sqrt{ (cL)^{-N_{LL}}N_{LL}!}} {  {\displaystyle \prod\limits_{j=1}^{N_{LL}/2} \frac{\lambda_j}{c}   \sqrt{\frac{\lambda_j^2}{c^2} + \frac{1}{4} } }  }  \sqrt{\frac{\det_{j,k=1}^{N_{LL}/2} \tilde{G}^{+}_{jk}} { \det_{j,k=1}^{N_{LL}/2}\tilde{G}_{jk}^- } }\epp
\end{equation}
The matrices $\tilde{G}_{jk}^{\pm}$ are similar to the Gaudin matrix $G_{jk}$ of the Lieb-Liniger model \cite{KorepinBOOK, 1981_Gaudin_PRD_23}, but with a different kernel:
\begin{equation}\label{eq:Gaudin_pm_LL}
\tilde{G}^\pm_{jk} = \delta_{jk} \Big( L + \sum_{l=1}^{N_{LL}/2} \tilde{K}^+(\lambda_{j},\lambda_{l}) \Big) - \tilde{K}^\pm (\lambda_{j},\lambda_{k}) \epc
\end{equation}
where $\tilde{K}^{\pm}(\lambda,\mu) = \tilde{K}(\lambda - \mu) \pm \tilde{K}(\lambda + \mu)$ and $\tilde{K}(\lambda) = 2c/(\lambda^2 + c^2)$. Hence we proved, starting from the XXZ off-shell formula (\ref{eq:overlap_XXZ_offshell}), the formula for overlaps of the BEC state with Bethe states of the Lieb-Liniger Bose gas \cite{LLpaper} for an arbitrary number of bosons $N_{LL}$. Note that in~\cite{LLpaper} the quotient of determinants is presented in a different way, but can be easily transformed into our representation using the relation 
\begin{equation}
	\det{}_{\!N}\left(\!\!\begin{array}{cc} A & B \\ B & A \end{array}\!\!\right) = \det{}_{\!N/2}(A+B)\det{}_{\!N/2}(A-B)
\end{equation}
for block matrices. Note furthermore that equation~\eqref{eq:Overlaps} holds for any solution of LL Bethe equations, irrespective of whether the Bethe roots are purely real numbers (as is the case in the repulsive regime $c>0$ of the Lieb-Liniger Bose gas) or form complex string solutions (which can occur in the attractive regime $c<0$).

\section{Summary}
In this paper we presented a rigorous proof of the BEC Lieb-Liniger overlap formula of~\cite{LLpaper} using the formula for overlaps of the N\'eel state with XXZ off-shell Bethe states, which was proven in~\cite{XXZpaper}. We sent parameters to infinity to recover global symmetry operators that act, in the scaling limit to Lieb-Liniger, on the N\'eel state as global $SU(2)$ operators. In this way the number of down spins could be reduced to a fixed finite number and the resulting state could be identified with the initial state of finitely many uniformly-distributed bosons. This allowed to gain the formula for overlaps of Lieb-Liniger Bethe states with this initial state, which has a nice application in the context of the KPZ equation  \cite{Calabrese_1402.1278} that is related to the attractive Lieb-Liniger Bose gas. Another nice application is the solution of the interaction quench to repulsive bosons in~\cite{LLpaper} using the so-called quench action approach \cite{2013_Caux_PRL_110}. 

Furthermore, using the results of~\cite{Pozsgay_1309.4593}, we related overlaps of $q$-raised N\'eel states to overlaps of different initial states which lie, as well as the $q$-raised N\'eel state itself, in a non-zero magnetization sector of the spin chain. This extends the results of~\cite{XXZpaper} where only the N\'eel state was considered.

The connection between the overlaps for the two different models, the XXZ spin chain and the Lieb-Liniger Bose gas, opens a way to discover more initial states for the Lieb-Liniger model. One could, for example, create non-uniform states by connecting the scaling limit to Lieb-Liniger with the limit of the rapidities which are sent to infinity. This can lead to different initial states and, of course, to different Gaudin-like determinant expressions, depending on how we send the rapidities to infinity. 

The leading behavior of these determinants in the thermodynamic limit can be evaluated which allows then for an exact analysis of non-equilibrium dynamics using the quench action approach proposed in~\cite{2013_Caux_PRL_110}. Within this method the time dependence of expectation values of certain operators is in principle accessible. Especially in the large time limit they can be expressed as expectation values of a single state, the so-called saddle point state. Since correlation functions for both models, the spin-1/2 XXZ chain and the Lieb-Liniger Bose gas, are related to each other \cite{2007_Seel, Pozsgay_JStatMech_P11017}, it would be interesting to investigate them regarding the saddle-point state. We will address these questions in an up-coming publication \cite{XXZpaper2}.

\vspace{3ex}

\section*{Acknowledgements}
I would like to express my gratitude to Frank G{\"o}hmann, Jean-S{\'e}bastien Caux, Jacopo De Nardis, and Bram Wouters for useful discussions. I thank Pasquale Calabrese for pointing out a sign mistake in equation~\eqref{eq:Gaudin_pm_LL}. I also thank the Netherlands Organisation for Scientific Research (NWO) for financial support.

\end{document}